\documentclass[%
 reprint,
 amsmath,amssymb,
 aps,
]{revtex4-2}
\usepackage{graphicx}
\usepackage[utf8]{inputenc}
\usepackage{listings}
\usepackage{parskip}
\usepackage{amsmath}
\usepackage[margin=1in]{geometry}

\usepackage{xcolor}

\usepackage{color}

\definecolor{ccomments}{rgb}{1,0.2,0.2}

\definecolor{cmissing}{rgb}{1,0.5,0.3}

\begin{document}


\title{Cell-induced wrinkling patterns on soft substrates}

\author{Aleksandra Arda\v{s}eva$^{1,2*}$, Varun Venkatesh$^{1*}$, Daiki Matsunaga$^{3}$, Shinji Deguchi$^{3}$, Amin Doostmohammadi$^{1 \dagger}$}
\address{$^{1}$ Niels Bohr Institute, University of Copenhagen, Denmark\\
  $^{2}$ Institute of Bioengineering, School of Life Sciences, \'{E}cole polytechnique f\'{e}d\'{e}rale de Lausanne (EPFL), Switzerland\\
 $^{3}$ Division of Bioengineering, Graduate School of Engineering Science, Osaka University, Japan\\
 $*$ These authors contributed equally.
$\dagger$ doostmohammadi@nbi.ku.dk.}

\begin{abstract}
 Cells exert traction forces on compliant substrates and can induce surface instabilities that appear as characteristic wrinkling patterns. Here, we develop a mechanical description of cell-induced wrinkling on soft substrates using a thin film elastic framework based on the F\"{o}ppl-von K\'{a}rm\'{a}n equations coupled to a phase-field model of a single cell. We model in-plane contractile stresses driven by cellular activity and study how their magnitude, spatial distribution, and symmetry determine the onset of wrinkling and the resulting pattern selection. The theory predicts transitions between distinct morphologies, such as radial, circumferential, and anisotropic wrinkle arrangements, and provides scaling relations for wrinkle wavelength and amplitude as functions of elastic parameters and imposed cellular forcing. We compare these predictions with available experimental observations of cell-driven wrinkling on compliant gels and find good agreement for both qualitative pattern classes and quantitative wavelength trends. Our results offer a minimal modelling framework to interpret wrinkling assays and connect observed surface patterns to underlying cellular forces.
\end{abstract}


\maketitle



\section{Introduction}

Mechanical interactions between cells and their environment are fundamental for cell-sensing and decision-making in both development and homeostasis. In tissues, cells rely on a delicate balance between internal pulling forces, dictated by cytoskeletal tensions, and external forces that arise from the microenvironment \cite{fletcher2010cell}. This mechanical interplay profoundly impacts protein distribution and gene expression within cells, inducing specific cellular functions \cite{panciera2017mechanobiology}. For instance, stretching the cell membrane results in enhanced cell division \cite{gudipaty2017mechanical} and directs differentiation pathways in stem cells \cite{vining2017mechanical}, whereas mechanical compression governs cell death and subsequent expulsion from tissues \cite{saw2017topological}.

Cells, however, are not only passive recipients of mechanical cues but actively respond to the environment and modify it by exerting contractile forces generated by cross-bridging interactions of actin and myosin filaments while moving \cite{discher2005tissue}. Recent advances in traction force microscopy (TFM) demonstrated that contractile forces cause rapid and long-ranged topographic anisotropies in the substrate, such as strains or wrinkles, which provide environmental cues and the means for substrate-mediated cell interactions \cite{winer2009non, han2018cell}. For instance, endothelial cells are capable of detecting and responding to substrate strains created by traction forces from their neighbours \cite{reinhart2008cell}.

Building on these insights, TFM measurements were recently combined with machine learning algorithms to introduce wrinkle-force microscopy (WFM) -- a method where cell-generated forces are revealed from the folding patterns in real-time by just imaging cells under the microscope, allowing the measurement of the wavelength and length of folds \cite{li2022wrinkle}. Despite this experimental progress, the quantitative relationship between active cell-generated forces and the emergent patterns of substrate folds remains poorly understood, limiting our ability to predict cellular behaviour from mechanical measurements.

The wrinkling phenomenon observed in cell-substrate systems is part of a broader class of mechanical instabilities that spans multiple biological scales and contexts. Wrinkling patterns emerge in systems ranging from leaf growth \cite{xiao2011modeling}, and shell buckling \cite{moulton2020mechanics} to algae development \cite{dumais2000whorl} and bacterial biofilm expansion \cite{geisel2022role}. This universality suggests that common physical principles govern pattern formation across diverse systems. Experimental studies of wrinkling in biological contexts that do theoretical analysis have predominantly only used controlled, idealized systems such as models of thin plates or shells \cite{jackson2023scaling}. Extensive theoretical work has characterized pattern formation and selection in response to various stress fields and boundary conditions \cite{li2012mechanics}. In particular, previous analytical efforts have established scaling laws~\cite{cerda2004geometry, huang_nonlinear_2005} and identified instabilities and transitions between different buckling modes \cite{cao_wrinkles_2011, jin_mechanics_2015}. Furthermore, numerical simulations have explored increasingly complex scenarios, from inhomogeneous substrates \cite{yang_wrinkle_2017, colin_layer_2019} on varied surface geometries \cite{tobasco2022exact, krause_wrinkling_2024} to hierarchically branched wrinkle structures \cite{ma_tunable_2020}. While these studies have provided fundamental insights into wrinkling mechanics, they have generally not addressed the specific challenges posed by biological systems, where forces are highly localized, temporally dynamic, and coupled to cellular responses.

F\"{o}ppl-von K\'{a}rm\'{a}n (FvK) setup for thin plate deformation is commonly used to study wrinkling patterns in compliant substrates \cite{Landau_book}. In the typical experimental setup, a bilayer thin film consists of a thin elastic layer residing on a thicker viscoelastic foundation. By applying constant compressive stress to the thin elastic layer, the evolution and morphology of wrinkles can be analysed systematically \cite{Huang_Im_2005, Huang_Im_2006}. These studies have successfully described wrinkling under controlled loading conditions, providing valuable insights into wavelength selection, pattern transitions, and stability criteria. However, much of previous work has focused on either isotropic or spatially and/or temporally uniform stress fields that cause wrinkling \cite{budday_wrinkling_2017} -- scenarios that differ from those encountered in cell-substrate interactions.

The traction forces exerted by cells not only vary dramatically between cell types \cite{aranson2016physical}, but are inherently dynamic, evolving as cells migrate, divide, or change shape. Moreover, cells actively sense and respond to not only the substrate topography \cite{li_micropatterned_2023}, but also regularly remodel their local environment \cite{pamonag_individual_2022} and have long been known to deform substrates \cite{harris_silicone_1980}. This creates a complex feedback loop: cell forces generate wrinkles, wrinkles influence cell behaviour, and modified cell behaviour produces new force distributions. The combined effects of these have recently been shown to produce orientational order for coarse-grained cell layers that otherwise would remain isotropic and stationary~\cite{venkatesh_nematic_2025}. The effects of a single cell, however, remain unknown.

To capture substrate-mediated cellular communication, previous work has modelled cells analytically as contractile force dipoles embedded in an elastic medium, demonstrating that elastic interactions can lead to attraction or repulsion between cells depending on their relative orientations \cite{schwarz2002elastic}. While this dipole approximation has proven useful for understanding long-range cell-cell interactions, such simplified models cannot capture the fine spatial details of force distributions nor the non-linear mechanics of substrate folding that characterize the cell-substrate feedback loop. Thus, there exists a need for a computational modelling framework that can accurately predict both the detailed wrinkling patterns and the underlying cell traction forces in a self-consistent manner.

In this paper, we present a computational framework for modelling wrinkling patterns caused by spatially heterogeneous, time-varying cell traction forces. Our approach combines the F\"oppl-von K\'{a}rm\'{a}n equations for thin elastic substrates with an active nematic phase-field model for a cell to capture the dynamic coupling between cellular forces and substrate deformation. We demonstrate correct prediction of wrinkling patterns and characterise the effects of the substrate stiffness, cellular aspect ratio, and cellular activity on the patterns quantitatively. Following this, we validate our model using experimental WFM data of fibroblast cells, demonstrating that it accurately reproduces observed wrinkle wavelengths and orientations not only by using experimental cell stresses but also with a coupled cell-substrate numerical model. 

\section{Methods}
\begin{figure}[htb!]
\centering
\includegraphics[width=1\linewidth]{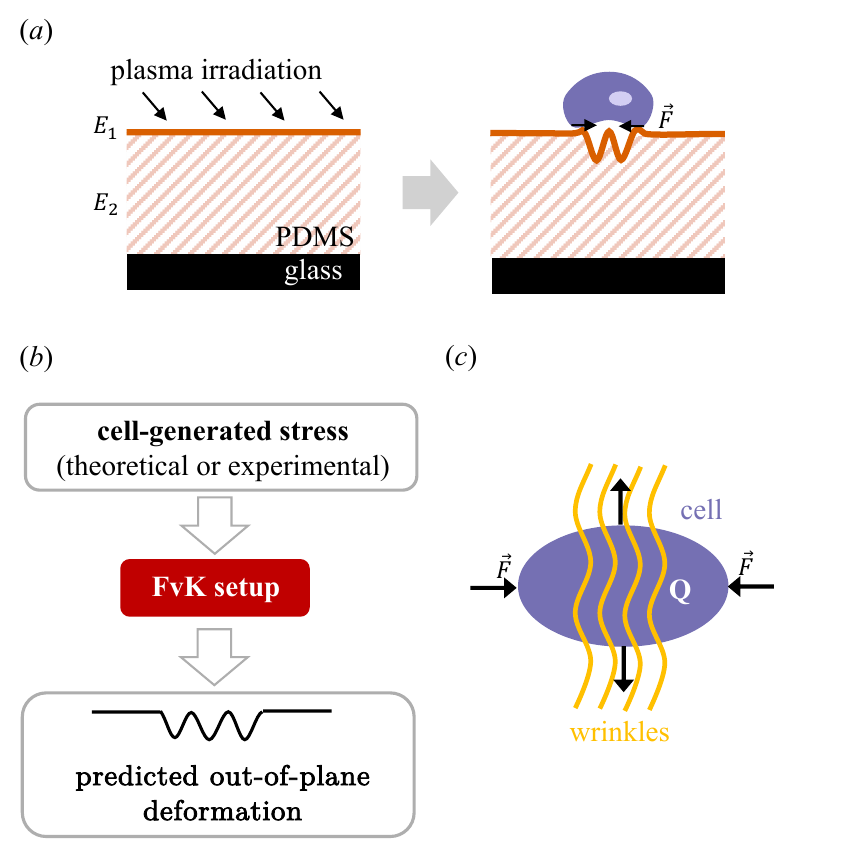}
\caption{\label{fig1} \textbf{Schematic representation of the modelling framework.} (a) Schematic representation of the experimental setup. PDMS is deposited on a glass surface. The top part is then plasma irradiated such that a thin layer of higher elasticity (Young modulus $E_1 > E_2$) is formed. The cell is placed on top of a stiffer layer, leading to a wrinkling pattern formation. (b) Simulation algorithm: we use a 2D phase-field model of a single cell to obtain the in-plane components of stresses that the cell exerts onto the substrate. The stress tensor is then used in the F\"{o}ppl-von K\'{a}rm\'{a}n setup to calculate the in- and out-of-plane deformation of this plate. (c) Schematic illustration of how the cell contractility is modelled.}
\end{figure}

In this paper, we model the wrinkling patterns in the experimental setup introduced by \cite{yokoyama2017new, li2022wrinkle}. The substrate consists of a thick layer of PDMS placed on a glass surface. The PDMS is then plasma irradiated, leading to the formation of a thin layer of higher elasticity (Fig. \ref{fig1}(a)), which can undergo deformation when a cell is placed on top. As the cell exerts contractile stresses, the substrate deforms, and wrinkling patterns appear. To model such a system, we couple the F\"{o}ppl-von K\'{a}rm\'{a}n equations of thin plate deformation with the 2D nematohydrodynamic model of a single cell that exerts traction forces onto the substrate (Fig. \ref{fig1}(b)).

\subsection{Modelling cell contractility}

There have been a number of studies, with different methods, that have used an active drop on a substrate to represent the dynamics of a single cell~\cite{joanny_drop_2012,  PhysRevLett.123.248006, shankar_optimal_2022}.  We employ the lyotropic nematic model to construct an active droplet \cite{blow2014biphasic, hughes_collective_2020, tiribocchi_crucial_2023} that captures the traction forces produced by the actin-myosin system in two dimensions. We represent the cell by defining a scalar field, $\phi\in[0,1]$, where $\phi=1$ corresponds to the nematic interior of the cells, whereas $\phi=0$ is the exterior medium (isotropic). We assume that the cell is stationary, hence its shape is governed by 
\begin{equation}
    \partial_t \phi = - \frac{\delta \mathcal{F}}{\delta \phi},
\end{equation}
where the free energy, $\mathcal{F} = \mathcal{F}^Q+\mathcal{F}^{CH}$, has contributions from the nematic, $\mathcal{F}^Q$, and the Cahn-Hilliard, $\mathcal{F}^{CH}$, energies. Cahn-Hilliard term sets the shape of the cell and the dynamics of the interface:
\begin{equation}
    \mathcal{F}^{CH} = \frac{\gamma}{\lambda}\int \textup{d}\vec{x} \left( 4\phi^2(1-\phi)^2+\lambda^2 (\vec{\nabla\phi})^2 \right).
\end{equation}
Here, $\lambda$ is the width of the diffusive interface and $\gamma$ is the surface tension. The first term defines a double-well potential with two minima: $\phi=1$ (inside) and $\phi=0$ (outside). Since we are only interested in the scenarios where cells are stationary, we prescribe cell shape by creating a binary field with $\phi(t=0)=1$ representing the cell and $\phi(t=0)=0$ being elsewhere. Using the above-described Cahn-Hillard free energy, the binary field is then relaxed for a short while to attain a finite gradient and non-zero thickness, essentially smoothing the field. This smooth phase field is consequently fixed and is not evolved, though the rest of the simulation. Hence, the cell boundary occurs at $\phi=0.5$.

In order to incorporate cell contractility, we assign a nematic tensor to the cell, $\textbf{\textup{Q}}=2(\vec{n}\vec{n} - \textbf{\textup{I}}/2)$, where $\vec{n}=(\cos \theta, \sin\theta)$ is the director and $\textbf{\textup{I}}$ is the identity tensor. The dynamics of $\mathbf{Q}$ is governed by the Beris-Edwards equation, which is coupled to the Navier-Stokes equations: 
\begin{align}
\partial_i v_i &= 0, \label{eq:NS_continuity}\\
\partial_t u_i& + u_j \partial_j u_i = \partial_j\sigma_{ij}^{\text{cell}} - \alpha u_i, \label{eq:NS_momentum} \\
D_t Q_{ij}& + S_{ij} = -\frac{\delta \mathcal{F^Q}}{\delta Q_{ij}}, \label{eq:Q_dynamics} 
\end{align}
where $D_t$ denotes the material derivative and $\mathbf{S}$ is the co-rotation term, defined as:
\begin{equation}
\begin{split}
    \mathbf{S} &= \left( \lambda \mathbf{E} + \boldsymbol{\Omega} \right) \left( \mathbf{Q} + \frac{\mathbf{I}}{2} \right)
    + \left( \mathbf{Q} + \frac{\mathbf{I}}{2} \right) \left( \lambda \mathbf{E} - \boldsymbol{\Omega} \right) \\
    &\quad - 2\lambda \left( \mathbf{Q} + \frac{\mathbf{I}}{2} \right) (\mathbf{Q} : \nabla \mathbf{u}),
\end{split}
\end{equation}
and $\frac{\delta \mathcal{F}}{\delta Q_{ij}}$ represents the functional derivative of the nematic free energy, $\mathcal{F}^{Q}$, given by:
\begin{equation}
\begin{split}
    \mathcal{F}^{Q} = \int \mathrm{d}\mathbf{x} \bigg[ & \frac{A}{2} \left(S_{nem}\phi - \frac{\mathbf{Q}^2}{2} \right)^2 \\
    & + \frac{k}{2} |\nabla \mathbf{Q}|^2 + \frac{L}{2} (\nabla \phi \cdot \mathbf{Q} \cdot \nabla \phi) \bigg],
\end{split}
\end{equation}
where the first term represents the bulk liquid crystal energy with $2S^2_{nem} = Q_{ij}Q_{ij}$, while the second and third terms penalise gradients in $\mathbf{Q}$ and the coupling to $\phi$, respectively.

Finally, the total cell stress, which is then passed into F\"{o}ppl-von K\'{a}rm\'{a}n setup, is calculated as $\sigma_{ij}^{cell} = \sigma^{fluid}+\sigma^{CH}+\sigma^{active}$. In particular, we include contributions from the fluid, the scalar phase field, and the dipolar activity. They are defined as
\begin{align}
\sigma_{ij}^{\text{fluid}} &= -p\delta_{ij} +  \eta (\partial_i u_j + \partial_j u_i), \\
\sigma_{ij}^{\text{CH}} &= \left( \mathcal{F}^{CH} - \frac{\delta F^{CH}}{\delta \phi}\phi \right) \delta_{ij} -\lambda \gamma (\partial_i \phi)(\partial_j \phi), \\
\sigma_{ij}^{\text{active}} &=   -\zeta (\delta_{ij} + Q_{ij}),
\end{align}
where $p$ is pressure, $\eta$ is viscosity, and $\zeta$ is the dipolar activity coefficient. Notably, we include an isotropic contribution to the activity  ($-\zeta \delta_{ij}$), which is usually disregarded in active nematics as this term drops out upon taking the gradient of active stress. 

The dominant contribution to the stress in our system arises from the activity, driving the majority of the dynamics. The sign of activity, $\zeta$, sets whether an ellipsoidal cell produces forces that contract or extend the substrate. Consider, for example, the following stress field with the director aligned with the x-axis:

\begin{equation}
    \mathbf{\sigma}^{active} = -\zeta\mathbf{Q} = - \zeta\begin{bmatrix}
1 & 0\\
0 & -1
\end{bmatrix} =  \zeta\begin{bmatrix}
-1 & 0\\
0 & 1
\end{bmatrix}.
\end{equation}
This active stress produces a force $\vec{F} = \vec{\nabla \phi} \cdot \sigma^{active}$ that is contractile (Fig.~\ref{fig1}(c)).

\subsection{Modelling the deformation of the substrate}
To model the wrinkling patterns caused by the cell traction forces, we closely follow the modelling approach from \cite{Huang_Im_2005, Huang_Im_2006} for evolving wrinkles in a bi-layer consisting of elastic and viscoelastic layers (top and bottom, respectively, in Fig. \ref{fig1}(a)).

We define deformations in $x$-, $y$-, and $z$-planes as $\omega_x$,  $\omega_y$, and $\omega_z$, respectively. $\omega_x$, $\omega_y$ are in-plane components, and $\omega_z$ is out-of-plane deformation. Assuming that the top layer is infinitesimally thin, the evolution of deformations is described using a thin plate von von K\'{a}rm\'{a}n theory \cite{Landau_book}, where the stress, $\sigma_{ij}$, and strain, $\epsilon_{ij}$, are defined as
\begin{align}
    \sigma_{ij} &= \sigma^{cell}_{ij} + 2\mu_f\epsilon_{ij} + \frac{\nu_f}{1-\nu_f}\delta_{ij},\\
    \epsilon_{ij} &= \frac{1}{2} \left( \frac{\partial \omega_i}{\partial x_j} + \frac{\partial \omega_j}{\partial x_i} \right) + \frac{1}{2} \left( \frac{\partial w_z}{\partial x_i} \frac{\partial w_z}{\partial x_j} \right).
\end{align}
Here, $\nu_f$ is the Poisson ratio,  $\mu_f$ is the shear modulus of the thin film, and $\sigma_{ij}^{cell}$ is the additional in-plane stress exerted by the cell that resides on top of the thin plate. 
We assume that $\mathbf{\sigma}^{cell}$ is symmetric. 

The system evolves according to the equations:
\begin{equation}
    \frac{\partial \omega_z}{\partial t} = -K \nabla^2 \nabla^2 \omega_z + F \nabla \cdot ( \boldsymbol{\sigma} \cdot \nabla \omega) -  \frac{\mu_R}{\eta_s} \omega_z,
\end{equation}
\begin{equation}
    \frac{\partial \omega_i}{\partial t} = \frac{Hh_f}{\eta} \nabla \cdot \boldsymbol{\sigma} -  \frac{\mu_R}{\eta_s} \omega_i,
\end{equation}
where $i = x, y$ denotes the in-plane displacements, and
\begin{align*}
    & K = \frac{(1-2\nu_s)\mu_f h_f^3 H}{12(1-\nu_s)(1-\nu_f)\eta_s},\\
    & F =  \frac{(1-2\nu_s)h_fH}{2(1-\nu_s)\eta_s}.
\end{align*}

\subsection{Numerical implementation}

We implement the modelling framework in an open-source software \texttt{OpenFOAM} \cite{weller1998tensorial}, which utilizes the finite-volume method for solving problems in fluid dynamics. In addition to the existing \texttt{OpenFOAM} fluid solver \texttt{PimpleFoam}, we have included the dynamics of the nematic tensor, $\textbf{Q}$, scalar phase field, $\phi$, and the deformable thin plate, $\omega_i$. The spatial resolution of the system is set to $\Delta x = 0.5$ with $L_x = L_y = 256$. In order to maintain the stability of FvK equations, we use different time resolutions for approximating the dynamics with one step corresponding to $\tau_f=1$ and $\tau_f=0.2$ for the dynamics of the fluid and thin plate, respectively. We initialise the system with random fluctuations of the order $Q, \omega \sim 10^{-3}$ with the velocity and pressure fields set to zero.  In the majority of our simulations, the phase field cell is initialised and fixed as an ellipse with a semi-major axis $a=55$ and a semi-minor axis $b=30$. 

The parameters used in the simulations are summarised in Table \ref{tab:parameters}. As stated in \cite{Huang_Im_2005}, the `rubbery modulus' is the elastic shear modulus at the rubbery limit. Since $H>>h$, we set $H=10$ and $h_f=0.5$. 
\begin{table}[!htb]
  \centering 
  \begin{tabular}{l|c|c}
\textbf{Parameter (Notation)} & \textbf{Value(s)} & \textbf{Dimensions} \\
    \hline
    Film shear modulus ($\mu_f$) & 0.1 & $ML^{-1}T^{-2}$ \\
    Substrate Poisson ratio ($\nu_s$) & 0.001 & $-$ \\
    Film Poisson ratio ($\nu_f$) & 0.01 & $-$ \\
    Substrate thickness ($H$) & 10 & $L$ \\
    Film thickness ($h_f$) & 0.5 & $L$ \\
    Substrate viscosity ($\eta_s$) & 10 & $ML^{-1}T^{-1}$ \\
    Rubbery modulus ($\mu_r$) & 0.001 & $ML^{-1}T^{-2}$ \\
    Cell interface width ($\lambda$) & 2.0 & $L$ \\
    Cell elasticity ($\gamma$) & 0.04 & $MT^{-2}$ \\
    Cell activity ($\zeta$) & 0.005--0.1 & $ML^{-1}T^{-2}$ \\
    Cell internal pressure ($\zeta_0$) & $\zeta$ & $ML^{-1}T^{-2}$ \\
    Cell aspect ratio ($\alpha$) & 0.2--1 & $-$ \\
    \end{tabular}
  \caption{Parameters/parameter ranges used in the numerical simulations.}
  \label{tab:parameters}
\end{table}

\section{Results}

We focus on modelling the wrinkling patterns in the substrate that arise from the traction forces exerted by a single stationary cell. We let the simulations run for one million time steps, until the system dynamics have reached a statistical steady-state. In particular, we investigate how different substrate and cell properties, such as substrate stiffness and cell shape, affect the resulting patterns. We then utilise stress fields and cell shapes obtained from wrinkle-force-microscopy \cite{li2022wrinkle} to reproduce the experimental patterns, thereby validating our model and predictions.

\subsection{Cell-induced wrinkling pattern formation}
\begin{figure}[!htb]
\centering
\includegraphics[width=\linewidth]{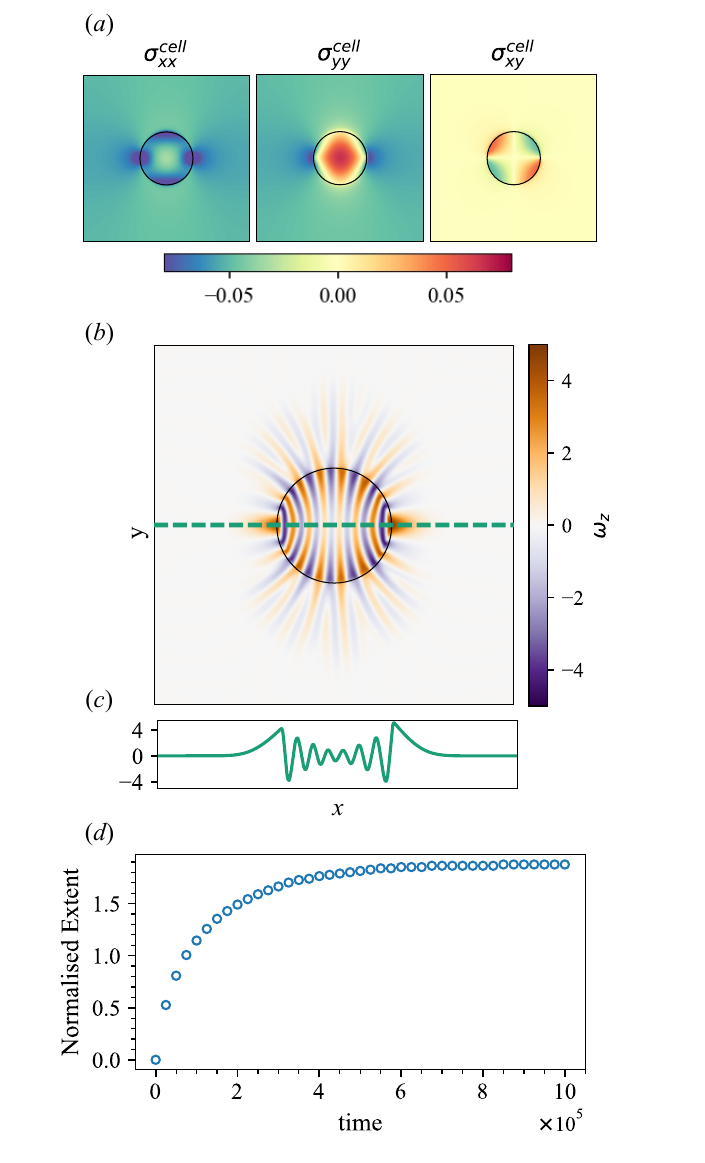}
\caption{\label{fig2} \textbf{Wrinkling pattern of a spherical cell.} (a) Snapshots of stress tensor components, $\sigma_{xx}^{cell}$, $\sigma_{yy}^{cell}$, $\sigma_{xy}^{cell}$, at the end of simulation. (b) Snapshot of a substrate height, $w_z$, obtained from the simulations. The solid black line represents the boundary of the cell. The green dotted line indicates a cross-section along which the height profile is extracted in panel (c). (c) Substrate height profile along the x-axis at $y=0$. (d) Normalised extent of wrinkles, i.e. distance the deformations propagate from the cell body, as a function of time, demonstrating that the system reaches values close to equilibrium.}
\end{figure}

Experimental observations highlight two key phenomena that we aim to reproduce in our simulations. First, wrinkles extend a considerable distance from the cell. Second, cells generate significantly more stress at their elongated ends compared to the short axis. As shown in Fig.~\ref{fig2}(a), a region of high stress concentration is visible along the elongated axis. This pattern arises from nematic anchoring, which confines the director field, leading to the requirement of a total topological charge of $+1$. However, in a 2D nematic system, this results in two $+1/2$ defects that make their way to the ends of the cell in order to minimise the elastic nematic energy. Since topological defects are known to generate significant stress, they are the primary cause of the high stress observed in this region. 

The localized stress at the cell poles drives the formation of wrinkles that propagate into the surrounding substrate. Fig.~\ref{fig2}(b) shows that these wrinkles extend well beyond the cell footprint and can even branch as they propagate. We have measured the distance the deformations propagate from the cell body (wrinkling extent) until a threshold of $\omega_z< 0.1$, as shown in Fig.~\ref{fig2}(d). The wrinkling extent increases rapidly during the initial stages before plateauing (Fig.~\ref{fig2}(d)), indicating that the system approaches a mechanical equilibrium and is able to develop the majority of its wrinkles on a relatively short time scale. Importantly, the deformation profile along the long axis reveals that wrinkle amplitude peaks near the cell boundary and decays both toward the cell interior and into the far field (Fig.~\ref{fig2}(c)). This gradual decay suggests that the wrinkles mediate mechanical communication over length scales much larger than the cell itself, motivating our focus on wrinkle extent in the analysis that follows.

\subsection{The effect of substrate properties}

Having obtained the expected wrinkle pattern for a single circular cell, we will now explore the effect of substrate properties on wrinkle formation. The wrinkle amplitude and wavelength in thin film systems have been well-characterized across many configurations, with established scaling laws relating these quantities to the elastic moduli and geometry of the bilayer \cite{Huang_Im_2006, ebata_wrinkling_2014, pal_faceted_2024}. While our specific single-cell scenario introduces unique features through the nematic-driven stress localization described above, the basic dependence of amplitude and wavelength on material properties follows known trends. Instead, we turn our attention to wrinkle extent since this quantity is particularly relevant for understanding mechanical signalling between cells. The distance over which wrinkles propagate defines an effective `sensing radius' within which neighbouring cells on the substrate can detect mechanical perturbations.
\begin{figure}[!htb]
    \centering
    \includegraphics[width=1\linewidth]{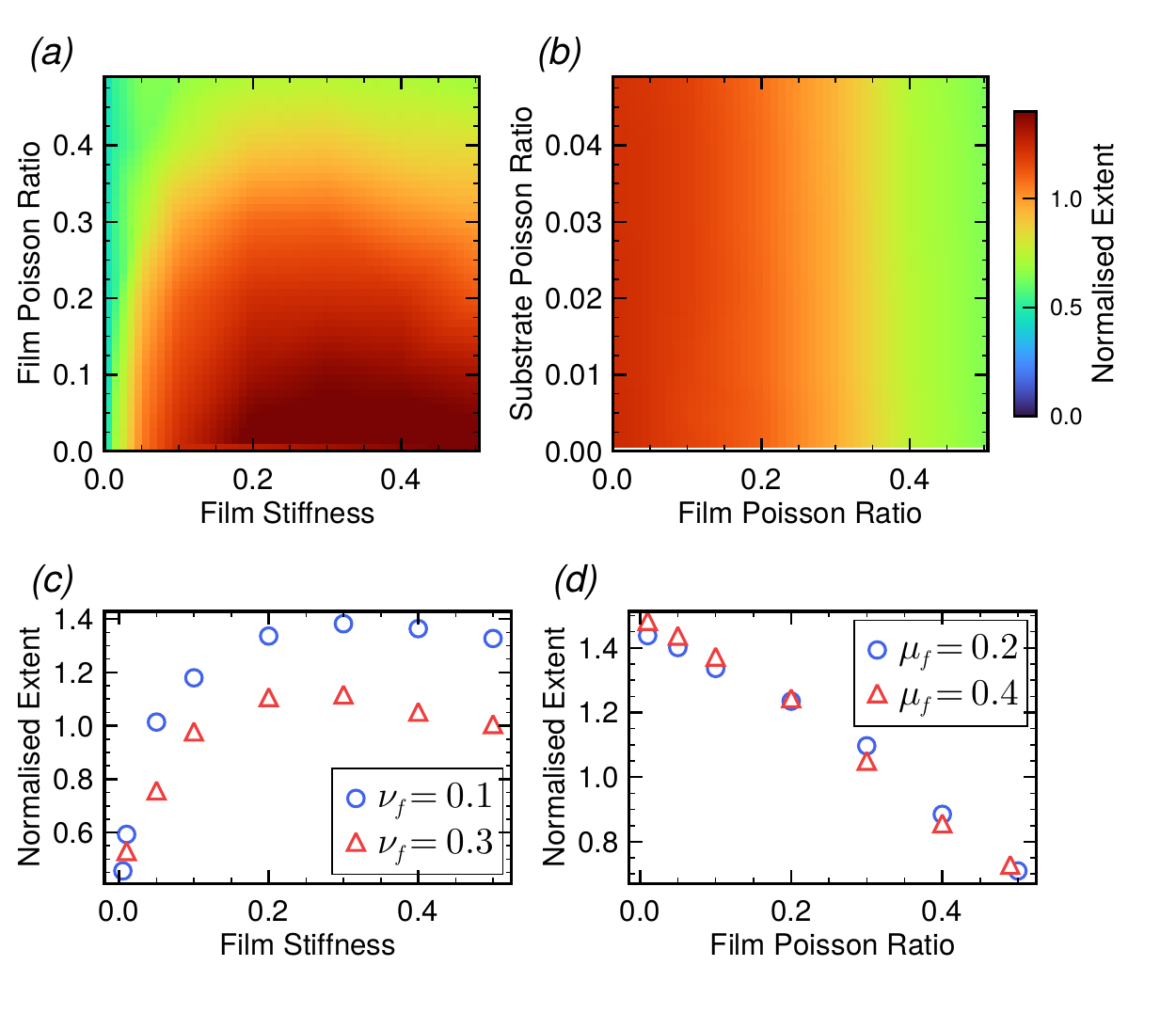}
    \caption{\textbf{Effect of material properties on the propagation of wrinkles from the cell.} Phase diagrams for $\mu_f$ vs $\nu_f$ (a), $\nu_s$ vs $\nu_f$ (b) characterise the propagation against experimentally-tunable parameters. The profile for the wrinkling extent as a function of $\mu_f$ keeping $\nu_f =0.01 $ and $nu_s = 0.001 $ (c) and against $\nu_f$ keeping $\mu_f = 0.1 $ and $nu_s = 0.001$ (d).}
    \label{fig:phase_diagrams}
\end{figure}
To characterize this behaviour systematically, we examine how wrinkle extent depends on the mechanical properties of both the film and the substrate. Fig.~\ref{fig:phase_diagrams}(a) shows the relationship between the film's shear modulus ($\mu_f$) and Poisson's ratio ($\nu_f$), while Fig.~\ref{fig:phase_diagrams}(b) explores the interplay between the compressibility of the two layers through their respective Poisson's ratios ($\nu_f$ and $\nu_s$) to show that while the wrinkling extent reduces linearly as a function of $\nu_f$ (Fig.~\ref{fig:phase_diagrams}(c)),  $\nu_s$ has minimal effect on propagation distance. In contrast, the dependence on $\mu_f$ exhibits a shallow maximum, beyond which the extent decreases slightly (Fig.~\ref{fig:phase_diagrams}(d)). These results demonstrate that the range of cell-generated deformations is sensitive to the film's elastic properties, particularly its compressibility and stiffness, while being relatively insensitive to the mechanical behaviour of the underlying substrate.

\subsection{Impact of cell shape on wrinkling pattern}
Having established how substrate properties modulate wrinkle propagation, we now examine the role of cellular properties on pattern formation. In particular, we begin by investigating the role of cell shape as this will affect the localisation of compressive stresses and, hence, the wrinkling patterns.

\begin{figure}[!htb]
\centering
\includegraphics[width=1\linewidth]{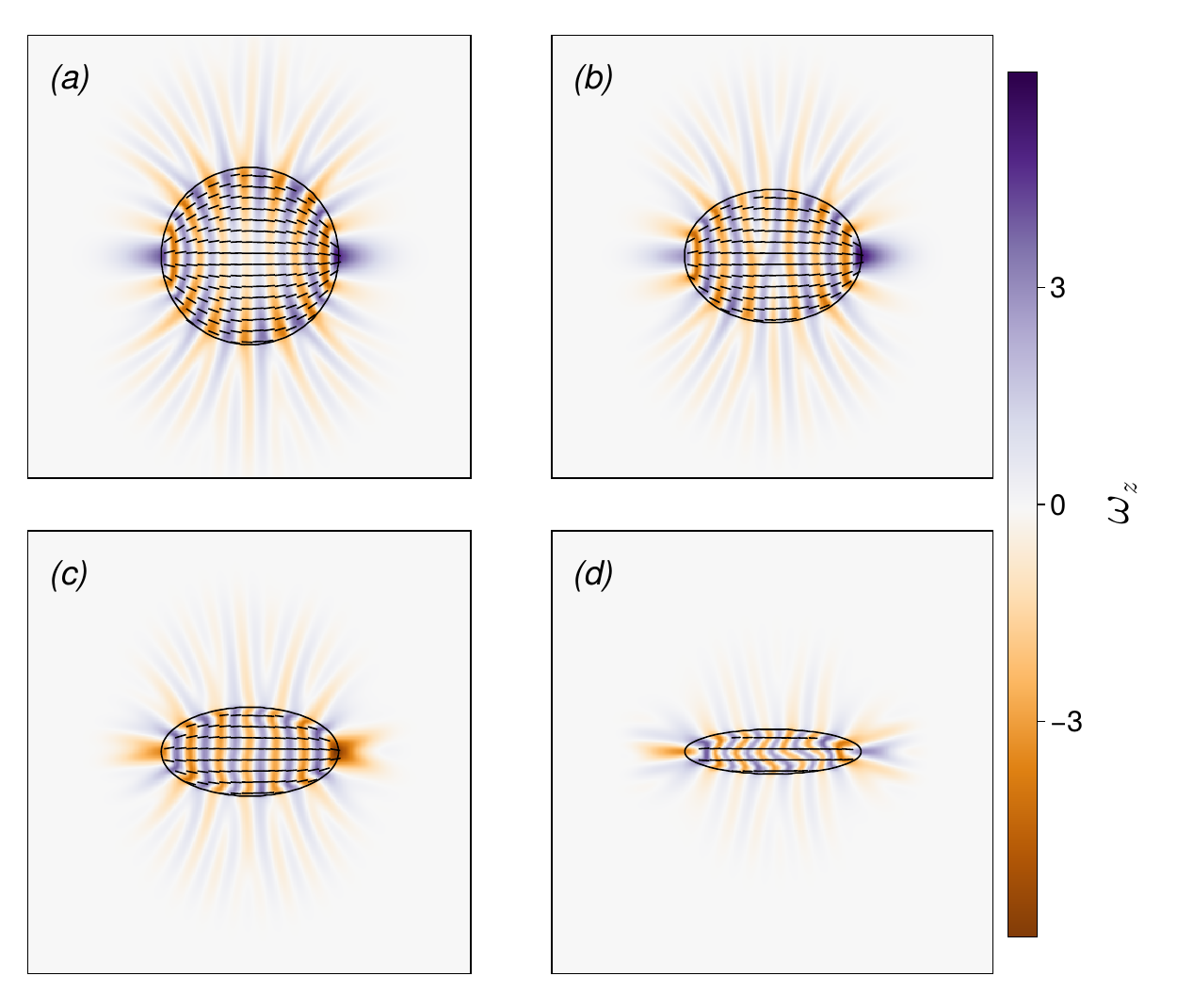}
\caption{\label{fig:cellshape} \textbf{Effect of cell aspect ratio on wrinkle patterns.} Plots of the wrinkle pattern as the aspect ratio is changed from an elongated ellipsoid shape to a circle (a--d). The colour map depicts the height of the wrinkle with the director field plotted as black lines. The contour of the phase field, $\phi = 0.5$, is drawn to show the outline of the cell. The activity in all is set to $\zeta = 0.04$.}
\end{figure}

To isolate this geometric effect, we vary the cell aspect ratio from circles to elongated ellipses while maintaining a contractile activity and fixed cell size via the major axis, as well as substrate properties. Fig.~\ref{fig:cellshape} shows representative wrinkling patterns for cells of varying aspect ratios. As cell shapes shift from circular to elongated, the extent of the wrinkles increases, with maximum extent achieved for completely circular cells. 

Changes in the cell shape affect not only the patterns outside of the cell boundary, but also wrinkled patterns underneath the cell itself. For ellipsoid cells, the wrinkles form a zig-zag pattern, whereas for circular cells, the wrinkles are straight and parallel to each other. The direct consequence of the sensitivity of wrinkle extent and pattern on even minimal change in the shape suggests that realistic cell shapes, which are affected by multitude of mechano-chemical factors \cite{phillip2021robust}, will serve as a powerful regulator of wrinkle patterns. 
At the same time, because cell shape is itself influenced by properties of the substrate \cite{yeung2005effects}, these results point to a feedback in which substrate mechanics shape cellular geometry, which in turn governs how the cell mechanically remodels its environment.

\subsection{Impact of cell contractility on wrinkling pattern}

Cell contractility, governed by actomyosin dynamics, represents a key control parameter that cells can actively regulate to modulate their mechanical interaction with the surrounding substrate \cite{panzetta_cell_2019}. To investigate this, we vary the activity parameter in our simulations and track the resulting changes in wrinkle characteristics. Specifically, we plot the mean amplitude and normalised wavelength in addition to the normalised wrinkle extent. The mean amplitude is defined as the spatial average of the field's magnitude, calculated over the region extending 10 units from the cell boundary. The wavelength is normalised by the major-axis length, where a value of $0.1$ indicates that there are 10 wrinkles within a cell. 
\begin{figure}[!htb]
    \centering
    \includegraphics[width=\linewidth]{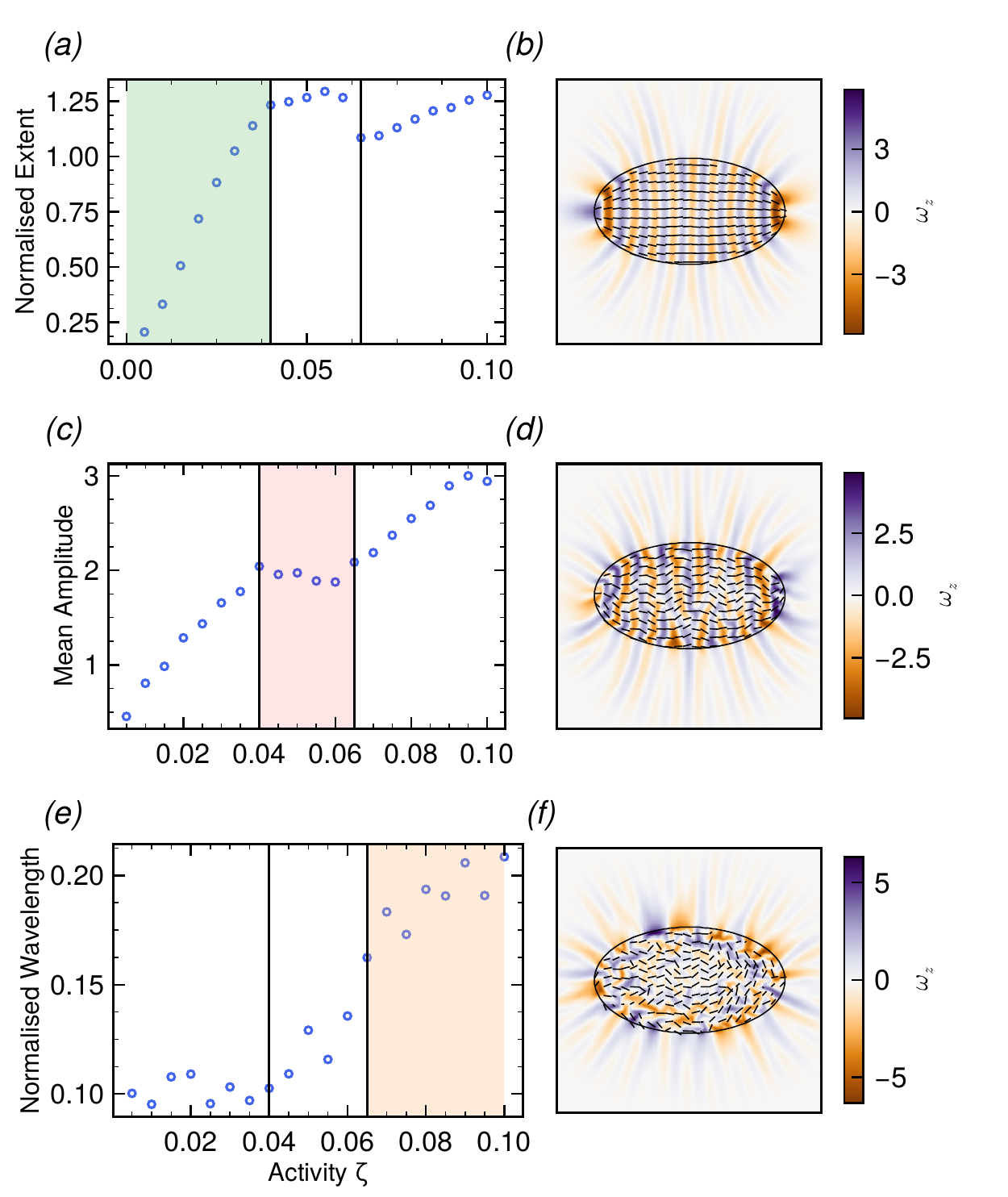}
    \caption{\textbf{Impact of cell contractility on wrinkling patterns.} Wrinkle properties, such as normalised extent (a), mean amplitude within 10 units from the cell boundary (c), and normalised wavelength (e) plotted against activity. Representative snapshots on the right display the director and wrinkle fields for different regimes in the nematic, corresponding to the marked vertical lines. These regimes correspond to a stationary nematic field (b), bend waves (d), and active turbulence (f). Both extent and wavelength are normalised by the major axis.}
    \label{fig:activescan}
\end{figure}

As shown in Fig.~\ref{fig:activescan}, the system exhibits three distinct dynamical regimes as contractility increases. At low activity levels, the system remains in a stationary nematic phase where the director field is static, and the stress distribution is time-invariant (Fig.~\ref{fig:activescan}(b)). In this regime, both wrinkle extent and amplitude grow in a controlled manner with increasing activity. The cell-substrate interaction is stable, allowing wrinkles to propagate efficiently into the substrate as contractile forces strengthen.

At intermediate activity levels, the system transitions into a regime characterized by bend wave propagation within the nematic field (Fig.~\ref{fig:activescan}(d)). This transition marks a qualitative change in behaviour for the wrinkle patterns. Both the amplitude and the extent of wrinkle propagation stagnate. The emergence of bend waves introduces temporal fluctuations in the stress field that appear to disrupt the coherent transmission of mechanical forces into the substrate, limiting how far wrinkles can extend despite the increasing cellular forces.

Further increases in contractility drive the system into an active turbulent regime, where the nematic field exhibits chaotic spatiotemporal dynamics (Fig.~\ref{fig:activescan}(f)). Interestingly, at the onset of this transition, wrinkle extent undergoes a drop before resuming its growth at higher activity levels. Amplitude, by contrast, continues to increase throughout this regime. The initial decrease in extent is likely caused by defects that are motile and thus can no longer provide sustained stress. However, as turbulent flow intensifies, the time-averaged stresses from not only the nematic but also the fluid become sufficiently large to drive wrinkle propagation despite the chaotic fluctuations. 

These results show that the increased cell contractility leads to larger and more spread out wrinkle patterns, both of which could be important in the context of cell-cell communication via the substrate. Moreover, we find that the specific organisation and relative time scale of the cell's internal organisation directly affect the resulting wrinkle pattern, which could be leveraged by cells to control the exact orientation of the pattern.

\subsection{Comparison with experimental data}
We next test our model predictions against experimental observations. By utilizing actual cell shapes and measured stress fields from wrinkle-force-microscopy experiments ~\cite{li2022wrinkle}, we directly compare simulated wrinkling patterns with those observed in living cells.

\begin{figure}[!htb]
\centering
\includegraphics[width=1\linewidth]{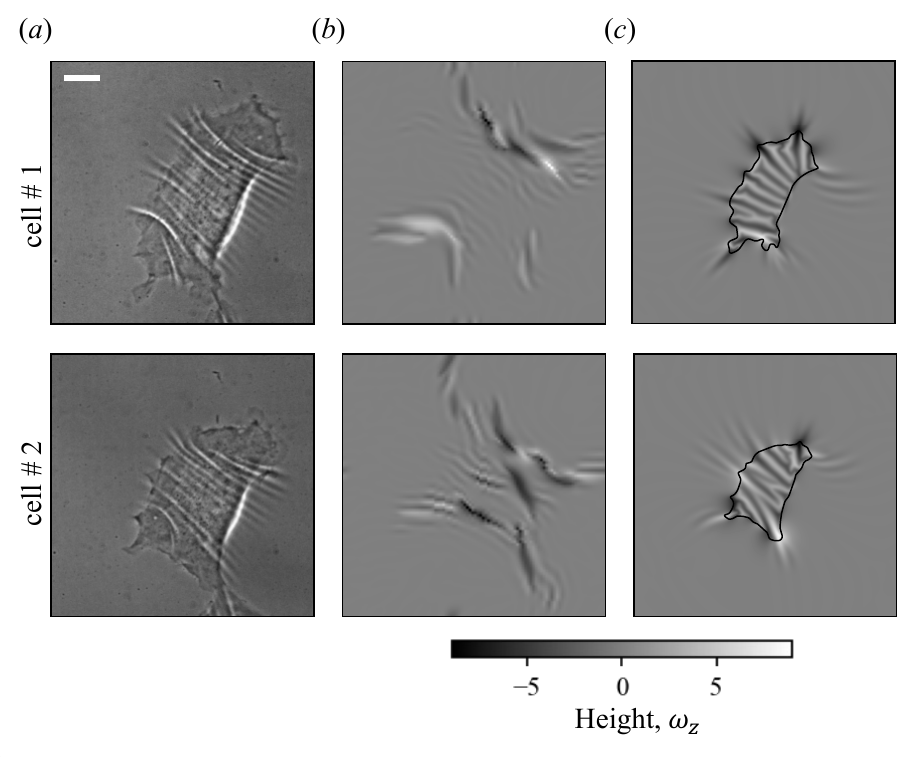}
\caption{\label{exp} \textbf{Model validation using experimental wrinkle-force-microscopy data.} (a) Experimental images of three cells and wrinkle patterns \cite{li2022wrinkle}. White scale bar in the top panel corresponds to 10 $\mu$m. (b) Substrate deformation along the z-axis, $\omega_z$, obtained by solving the FvK equations with stresses extracted from the experiments. (c) Wrinkling pattern obtained from the cell--FvK framework, where only cell shape is used as an input for the model. The black line represents cell contour (i.e. $\phi=0.5$). }
\end{figure}

In particular, we work with a dataset consisting of two cells, which are positioned away from the image boundaries to minimize edge effects, while still exhibiting clear wrinkling patterns. This prevents any errors arising from the sensitivity to boundary conditions. Each dataset provides an image of a single adherent cell and the corresponding substrate wrinkle pattern (Fig.~\ref{exp}(a)). For each of the cells, we extract the stress tensor from the experimental force measurements \cite{monfared2023mechanical}. These are then used as input, $\sigma_{ij}^{cell}$, to our mechanical model.

We solve the FvK equations using the experimentally measured stress as input with either periodic or Dirichlet boundary conditions. The obtained wrinkling patterns show good qualitative agreement with the experimental observation (Fig.~\ref{exp}(b)). Particularly, the alignment, spatial extent, and in some cases even the number of wrinkles match the experimental pattern. This qualitative agreement demonstrates the validity of our framework.

To demonstrate the predictive capability of the coupled cell-substrate framework, we performed simulations using only the cell shape as an input into the model, where the dynamics of active nematics generate stress distributions for the complicated shape of the realistic cell. The resulting wrinkle patterns again capture the essential features observed experientially (Fig.~\ref{exp}(c)).

The remaining differences between simulations and experiments can be attributed to a combination of experimental and modelling factors. On the experimental side, the finite field of view and imperfect knowledge of boundary conditions can influence the inferred deformation field. In addition, stresses reconstructed from traction force microscopy carry measurement uncertainty, which propagates into the wrinkling calculation. Despite these limitations, the model captures the main qualitative features across all three cases, and gives particularly close agreement for cells 1 and 2 in Fig. \ref{exp}(b). Notably, when we use cell shape alone and generate stresses consistently with the active nematic description, the resulting patterns retain the experimentally observed orientation and spatial organisation, including for the case where the stress-based reconstruction deviates most. This indicates that the coupled framework can be used as a forward model to connect readily observable cell geometry and wrinkle morphology to underlying contractile forcing, and can complement traction-based reconstructions when these are uncertain. 

\section{Conclusions}
We develop a modelling framework for cell contractility-induced wrinkling in soft, deformable substrates. By coupling nematic director dynamics of a cell to the F\"{o}ppl-von K\'{a}rm\'{a}n equations of thin plate deformation, our simulations reproduce key experimental features, including stress localization at cell poles and long-range wrinkle propagation in the substrate, which is important for intercellular signalling \cite{alisafaei_long-range_2021}. 

Sensitivity analyses of the substrate properties for the extent of wrinkle propagation showed that the important control parameters are the stiffness and Poisson's ratio for the irradiated film. Increasing film stiffness is helpful for substrate-based communication, but only until a certain amount. This finding might have implications beyond wrinkling, as substrate stiffness is a fundamental feature of cell-substrate interaction and contributes to everything from cell polarisation to self-propulsion \cite{ladoux_frontrear_2016,balcioglu_subtle_2020}. While properties of the substrate are important, we find that the cell itself plays an important and controlling part in the wrinkling process as well. First, cell geometry is seen to play a significant role, with circular cells producing maximum wrinkle extent compared to elongated shapes. Second, cell contractility exhibits three distinct dynamical regimes: (i) a stationary nematic phase with controlled wrinkle growth; (ii) a bend wave regime where extent stagnates despite increasing forces; and (iii) an active turbulent phase where wrinkles again grow. 

Finally, we compare our modelling against experimental data from wrinkle force microscopy, confirming that our model accurately reproduces observed wrinkling patterns when supplied with measured stress fields or cell shapes alone. Hence, the framework presented here establishes a direct link between single-cell dynamics and tissue-scale interactions, providing predictions for future \textit{in vitro} experiments. By defining wrinkle extent as an effective sensing radius, our work provides quantitative insight into how cells may mechanically communicate with their environment and potentially coordinate collective behaviours through substrate-mediated forces. 

Future extensions of this model should address multiple interacting cells \cite{monfared2026multiphase}, dynamic processes such as cell migration and shape changes \cite{bodor2020cell, driscoll2012cell}, and the role of substrate heterogeneity \cite{sunyer2016collective}. Such developments will be essential for understanding mechanical signalling in complex multicellular environments and bridging the gap between single-cell mechanics and tissue-level organization.

\subsection*{Acknowledgements} 
It is a pleasure to acknowledge helpful conversations with Alain Goriely and Dominic Vella at the early stage of this project. A.A. acknowledges support from the EU’s Horizon Europe research and innovation program under the Marie Sklodowska-Curie grant agreement No. 101063870 (TopCellComm) and SNSF Ambizione grant No. 232854. A. D. acknowledges funding from the Novo Nordisk Foundation (grant No. NNF18SA0035142 and NERD grant No. NNF21OC0068687), Villum Fonden (Grant no. 29476), and the European Union (ERC, PhysCoMeT, 101041418). Views and opinions expressed are, however, those of the authors only and do not necessarily reflect those of the European Union or the European Research Council. Neither the European Union nor the granting authority can be held responsible for them.

\subsection*{Author contributions} 
A.D. designed the project. V.V. implemented the modelling framework. A.A. and V.V. performed numerical simulations and analyses. A.A. and V.V. prepared the first draft of the manuscript. D.M. and Sh.D. conducted and provided the experimental data. All authors contributed to the writing of the manuscript.

\bibliographystyle{ieeetr}
\bibliography{refs}
\end{document}